\documentclass[useAMS,usenatbib]{mn2e}
\usepackage{latexsym,graphicx,natbib}

\newcommand\kms{{\rm\,km\,s^{-1}}}

\newcommand\msun{\rm\,M_\odot}
\newcommand\rsun{\rm\,R_\odot}

\title[Ejection of Hyper-Velocity Stars by Intermediate-Mass Black Holes]
      { Ejection of Hyper-Velocity Stars from the Galactic Centre 
	by Intermediate-Mass Black Holes}
      
\author[H. Baumgardt, A. Gualandris and S. Portegies Zwart]
{
  H. Baumgardt$^{1}$\thanks{e-mail: holger@astro.uni-bonn.de (HB);
    alessiag@science.uva.nl (AG); spz@science.uva.nl (SPZ)}, 
  A. Gualandris$^{2, 3}$\footnotemark[1] and
  S. Portegies Zwart$^{2, 3}$\footnotemark[1]\\ 
  $^{1}$Argelander Institute for Astronomy, University of Bonn, Auf dem H\"ugel 71, 53121 Bonn,
  Germany\\
	       $^{2}$Astronomical Institute ``Anton Pannekoek,'', University of
  Amsterdam, Kruislaan 403, Netherlands\\
  $^{3}$Section Computational Science, University of Amsterdam,
  Kruislaan 403, Netherlands\\
}

\begin{document}

\date{Accepted ????. Received ?????; in original form ?????}

\pagerange{\pageref{firstpage}--\pageref{lastpage}} \pubyear{2006}

\maketitle

\label{firstpage}

\begin{abstract}
We have performed $N$-body simulations of the formation of
hyper-velocity stars (HVS) in the centre of the Milky Way due to
inspiralling intermediate-mass black holes (IMBHs).  We considered
IMBHs of different masses, all starting from circular orbits at an
initial distance of 0.1\,pc.
We find that the IMBHs sink to the centre of the Galaxy due to dynamical friction,
where they deplete the central cusp of stars. Some of these stars
become HVS and are ejected with velocities sufficiently high to escape the Galaxy.
Since the HVS carry with them information about their origin,
in particular in the moment of ejection, the velocity distribution and
the direction in which they escape the Galaxy,
detecting a population of HVS will provide insight
in the ejection processes and could therefore provide indirect evidence for the existence of IMBHs.

Our simulations show that HVS are generated in short bursts which last only a few Myrs
until the IMBH is swallowed by the supermassive black hole (SMBH). 
HVS are ejected almost isotropically, which makes IMBH induced ejections hard to distinguish 
from ejections due to encounters of stellar binaries with a SMBH. 
After the HVS have reached the galactic halo, their escape velocities correlate 
with the distance from the Galactic centre in the sense that the fastest HVS can be found furthest away from 
the centre. The velocity distribution of HVS generated by inspiralling IMBHs is also nearly independent of 
the mass of the IMBH and can be quite distinct from one generated by binary encounters.
Finally, our simulations show that the presence of an
IMBH in the Galactic centre changes the stellar density distribution inside $r<0.02$\,pc 
into a core profile, which takes at least 100 Myrs to replenish.
\end{abstract}


\section{Introduction}
\label{sec:intro}

Hills (1988) was the first to show that the ejection of stars 
with velocities $>1000\kms$ is a natural consequence of galaxies 
hosting supermassive black holes. 
He named these stars 'hyper-velocity stars' (HVS). 
Recently, several HVS have been discovered in the Galactic
halo \citep{b05, hi05, b06a, b06b}.
Except for one star which might have been ejected from the LMC 
\citep{edel05}, the travel times of all HVS are short enough 
that the stars could have been ejected from the Galactic centre 
within the lifetimes of the stars, confirming Hills' predictions.

The exact formation mechanism of HVS is however still a matter of debate. 
The ejection of stars by supernova explosions in close binary systems 
\citep{b61} and dynamical encounters \citep{p67}
cannot produce main-sequence stars with velocities exceeding a few hundred 
kilometers per second \citep{gps05}, leaving the interaction of stars 
in galactic nuclei around super-massive black holes (SMBHs) 
as the only possible source for HVS.

Yu \& Tremaine (2003) considered three processes which could eject 
stars from the vicinity of SMBHs: (1) close encounters between two single stars, 
(2) encounters between stellar binaries and the central SMBH
and (3) encounters between single stars and a massive black hole binary.
In the case of the SMBH in the Galactic centre, they found that 
close encounters between single stars can eject stars with a rate 
of only $10^{-11}$\,yr$^{-1}$ which would create less than one HVS 
during the lifetime of the Milky Way. 
The other two processes were found to eject stars with rates of up to 
$10^{-4}$\,yr$^{-1}$, sufficiently high to explain the observed number of HVS 
in the halo of the Milky Way.

Similar results were also obtained by \citet{gps05},
who studied the ejection of stars by SMBHs by means of scattering experiments 
and found that HVS formation is possible by both processes. 
The tidal disruption of binaries was found to create HVS 
with higher velocities while the ejection from binary black hole 
systems creates HVS with a higher rate. Similarly, \citet{gl06} found
that the tidal breakup of stellar binaries can create HVS with velocities
significantly higher than what has been found so far.

A distinction between the two scenarios might come from a detailed analysis 
of the spatial and kinematical distribution of HVS:
the ejection of stars due to the interaction of stellar binaries with an SMBH 
should be nearly constant with time since the reservoir of binary stars 
in the Galactic centre is depleted only slowly. Furthermore, the distribution 
of binary orbits should be nearly isotropic in sufficiently relaxed nuclei, 
implying that the resulting HVS distribution will also be isotropic.
In contrast, the ejection of stars from an SMBH-IMBH binary should show 
characteristic variations with time and spatial direction: 
HVS are mainly ejected when the inspiralling IMBH reaches the centre 
since the density of stars is highest in the centre and the velocity 
dispersion is also highest close to the SMBH. In addition, escaping stars 
acquire their extra velocities mainly in the direction of motion 
of the IMBH, introducing a spatial anisotropy in the HVS distribution.

These considerations were confirmed by \citet{l05}, who studied analytically 
the distribution of escapers created by an SMBH-IMBH pair 
in a dense stellar cusp. 
He found that the ejection of stars in case of a black hole binary 
is happening mainly in a burst which lasts a few dynamical friction 
timescales and  that if the IMBH is initially in a nearly circular orbit, 
the velocity vectors of the ejected stars also cluster around 
the orbital plane. If the IMBH moves in an eccentric orbit, 
stars are ejected in a broad jet roughly perpendicular to the Runge-Lenz 
vector of the IMBHs orbit. Similarly, \citet{shm06} found through three-body
scattering experiments that eccentric black hole binaries eject stars along a
broad jet perpendicular to the semimajor axis of the binary.

IMBHs could form in galaxies through runaway collisions of stars 
in star clusters, giving rise to ultra-luminous X-ray sources 
\citep{pbhmm04}. They could later be brought into the centres of galaxies 
through dynamical friction of the star clusters \citep{pbmmhe06}.
Their merger with the central SMBHs would be an important source 
of gravitational waves, detectable with the next generation 
of gravitational wave detectors, like e.g. {\it LISA}. 
IMBHs might also be an important contribution
for the growth of SMBHs in the early universe \citep{ebi01}.

In the present paper we have therefore performed collisional 
$N$-body simulations 
of the dynamics of inspiralling IMBHs in stellar cusps around 
supermassive black holes. The aim of our simulations is to study whether 
the ejection of stars by IMBHs leads to observable consequences 
in the distribution of HVS which might help to distinguish between 
different ejection scenarios and which could give an indirect hint for the
presence of one or more IMBHs in the centre of the Milky Way.

\section{Description of the $N$-body runs}
\label{sec:Nbody}

All runs were performed with the collisional $N$-body code NBODY4
\citep{a99} on the GRAPE6 computers \citep{mfkn03} of Bonn 
and Tokyo University. Our runs contained three different components: 
a central super-massive black hole (SMBH), an IMBH and $10^5$ stars. 
In all simulations the SMBH was initially at rest at the origin 
and had a mass of $M_{\rm SMBH} = 3 \cdot 10^6\msun$, 
similar to the mass of the SMBH at the Galactic centre 
\citep{sch05, ghe05}.
The mass of the IMBHs was varied in the different runs.
In total we performed 3 runs, using IMBH masses of 
$M_{\rm IMBH}=10^3\msun$, $3 \cdot 10^3\msun$ and $10^4\msun$ 
respectively. All IMBHs moved initially in circular orbits at a distance 
of 0.1\,pc from the SMBH.

The stars had masses of $m_*=30\msun$ and were initially
distributed in a cusp around the SMBHs according to the following density law: 
\begin{equation}
 \rho(r) = \frac{\rho_0}{r^{7/4} \left(1 + \left(\frac{r}{r_0}\right)^5\right)}
\end{equation}
with $\rho_0 = 3 \times 10^5\msun\,{\rm pc}^{-5/4}$ and $r_0= 1$\,pc. 
For distances $r \ll r_0$, this distribution corresponds to 
a $r^{-7/4}$ power-law cusp which theoretical arguments 
and $N$-body simulations have shown to evolve in a stellar 
system around an SMBH \citep{bw76, baum04a, baum04b, pms04}. 
The chosen form  for the cusp is also compatible with
observations of the stellar density distribution in the Galactic centre, 
which shows a power-law cusp inside 10'' \citep{gen03}. 
It has the additional advantage that at larger distances, where the stellar 
distribution has little influence on the outcome of the simulations, 
the stellar density drops off quickly. The overall density was chosen 
in such a way to be compatible with current limits on the density of 
stars in the Galactic centre \citep{gen03}. 

Stars were merged with the black holes (both SMBH or IMBH) if their separation became 
smaller than their tidal radius given by
\begin{equation}
 r_t = \left(\frac{M_{\rm BH}}{m_*}\right)^{1/3} R_*.
\end{equation}
where $R_*$ is the radius of the stars, which we set equal to $1\rsun$.
The mass of disrupted stars was added to the mass of the black holes. 
IMBHs were merged with the central SMBH if they passed within 
the radius of the last stable orbit, assumed to be 3 Schwarzschild radii. 
We did not include the effects of gravitational radiation
into our runs. Neither was any softening used in calculating the gravitational
forces between the particles in our calculation.

Simulations were stopped when an IMBH merged with the SMBH or the runs reached 
15\,Myrs, whichever happened first.

\subsection{Scaling issues}

The mass of the ''stars'' in our simulations are probably too high 
compared to real Galactic nuclei, since even if the central parts are 
enriched in black holes which spiralled into the centres 
through dynamical friction (\citet{baum04b}, \citet{frb06}), 
the average mass is 
unlikely to be higher than a few $\msun$. 
We therefore have to study to which extent the high stellar masses in our runs 
can bias our results.

The inspiral time of the IMBHs should not change since dynamical friction 
is independent of the mass of the background particles as long as 
$M_{\rm IMBH} \gg m_*$ \citep{bt87}, which is the case both in 
our simulations and in real Galactic nuclei. Stochastic changes to the orbits 
of the IMBHs happen on a relaxation time scale and should scale with the average 
stellar mass $\langle m \rangle$ of cusp stars as $\langle m \rangle ^{-1}$, 
i.e. in real nuclei the IMBHs change their orbital parameters 
slower than  in our simulations, although the change should not 
be larger than a factor of a few.

The number of stars removed from the cusps after the IMBHs have spiralled 
into the centre will to first order  scale linearly with the number of stars 
present if the overall mass density profile is fixed, i.e. would roughly 
be a factor of 10 higher in the Milky Way than in our simulations. 
At later stages, the inner cusps become depleted and stars have 
to be scattered into low-angular momentum orbits through relaxation 
processes in the outer cusp. The number of stars scattered into 
low-angular momentum orbits scales as $dN \sim n(r)/t_r\,dt$, 
where $n(r)$ is the number density of stars at radius $r$ and 
$t_r$ is the relaxation time at radius $r$. Since the relaxation time scales 
with the mass of the stars as $\langle m \rangle ^{-1}$ \citep{s87},
and since the number of stars $n(r)$ scales as 
$\langle m \rangle ^{-1}$ if the overall mass density profile is fixed, 
the number of stars scattered into low-angular momentum orbits is to first 
approximation independent of the average mass of stars. We will come back
to this point in sec. \ref{sec:scal}.

\section{Results}
\label{sec:results}

\subsection{IMBH inspiral and core formation}

Initially all IMBHs sink towards the central SMBH due to dynamical friction. 
Since in all runs the masses of the IMBHs are much
higher than the masses of the background stars, the frictional drag 
on the IMBHs is given by (see \citet{bt87} Eq.\,7-18):
\begin{equation}
 \frac{d\vec{v}}{dt} = -\frac{-4 \pi \ln{\Lambda} G^2 \rho(r) M_{\rm IMBH}}{v^3} 
\left[ {\rm erf}(X) - \frac{2 X}{\sqrt{\pi}} e^{-X^2} \right] \vec{v}
 \label{dynf}
\end{equation}
where $\rho(r)$ is the background density of stars, 
$\ln{\Lambda}$ the Coulomb logarithm, and $X=\vec{v}/(\sqrt{2}\sigma)$
is the ratio between the velocity of the IMBH and the (1D) stellar 
velocity dispersion $\sigma$. For an $r^{-1.75}$ cusp profile,
$\sigma$ is approximately 0.60 times the circular velocity. 
Assuming that the IMBH moves in a circular orbit, and setting 
$\ln{\Lambda}=6.6$ \citep{sp03}, Eq.\,\ref{dynf} can 
be rewritten as:
\begin{equation}
 F = - 47.51 \frac{G^2 \rho_0 M^2_{\rm IMBH}}{v_c^2\,r^{1.75}}\,.
\end{equation}
Since the rate of angular momentum change is equal to $dL/dt=F r/M_{\rm IMBH}$ 
and since the angular momentum itself is given by 
$L = r v_c = \sqrt{G M_{\rm SMBH} r}$, 
this can be rewritten as:
\begin{equation}
 r^{-3/4}\,\frac{dr}{dt} = -95.0\frac{\sqrt{G} \rho_0 M_{\rm IMBH}}
{M_{\rm SMBH}^{3/2}}\,.
\end{equation}
Solving this equation with the initial condition $r_0 = 0.1$\,pc
gives for the radius reached at time $t$:
\begin{equation}
 r(t) = \left(r^{1/4}_0-23.8\frac{\sqrt{G} \rho_0 M_{\rm IMBH}}
{M_{\rm SMBH}^{3/2}} \; t\right)^4
\label{insp}
\end{equation}
and for the time required to reach the centre of the galaxy:
\begin{eqnarray}
\nonumber t_{\rm fric} & = & 0.0421 \frac{r_0^{1/4} M_{\rm SMBH}^{3/2}}
{\sqrt{G}\rho_0 M_{\rm IMBH}} \\
  & = & 6.11 \left( \frac{M_{\rm SMBH}}{3 \cdot 10^6\msun} \right)^{1.5} \left( \frac{M_{\rm IMBH}}{10^3\msun} \right)^{-1} \mbox{Myrs}\,.
\label{tfric}  
\end{eqnarray}

\begin{figure}
\begin{center}
\includegraphics[width=8.3cm]{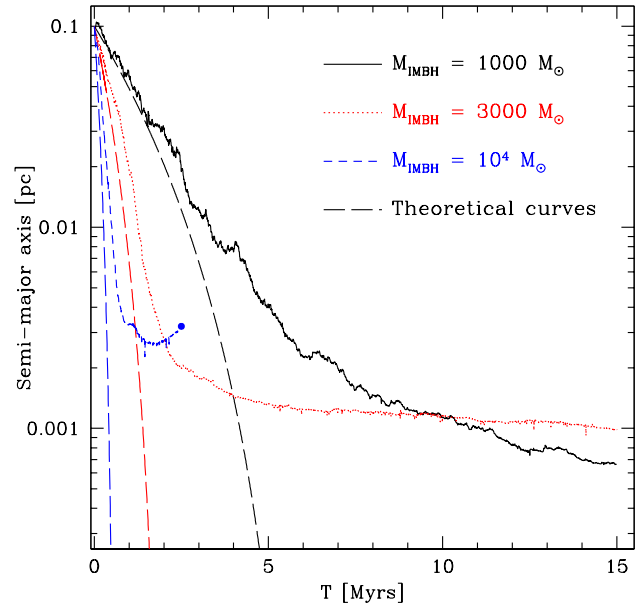}
\end{center}
\caption{Semi-major axis of the IMBHs as a function of time for 
the three runs performed. Solid lines show the measured semi-major axis, 
dashed lines the prediction from Eqs.\,\ref{insp} and \ref{tfric}. 
They agree quite well for the initial inspiral phase. 
At a distance of about $r \approx 0.003$\,pc, the inspiral of the IMBHs 
slows down in our simulations because the central cusps 
run out of stars.}
\label{fig:inspiral}
\end{figure}
Fig.\,\ref{fig:inspiral} compares the inspiral predicted by the above 
theory with the results of the $N$-body runs. We obtain reasonable agreement 
with the $N$-body data as long as the distance of the IMBHs is larger 
than $r>0.003$\,pc: the differences between the theoretical curves and 
the data are within the errors with which the Coulomb logarithm was determined by
\citet{sp03}.

Inside $r=0.003$\,pc the stellar cusps contain only few stars, 
(for the cusp profile chosen in our runs only 
$M(< \nolinebreak r)=2 \cdot 10^3 M_\odot$ in stars were inside this radius),
so the mass in stars becomes comparable to the mass of the inspiralling IMBHs. 
In this case dynamical friction becomes inefficient and the inspiral stalls 
since the IMBHs cannot displace enough stars to lose further orbital energy. 
The remaining evolution until the simulations were 
stopped at $T=15$\,Myrs is mainly driven by relaxation between stars 
at larger radii, due to which some stars are scattered into 
the central loss cone around the SMBH. The resulting interactions 
between the stars and the IMBHs cause a much slower inspiral of the IMBHs
\citep{bbr80,m97}.

\begin{figure}
\begin{center}
\includegraphics[width=8.3cm]{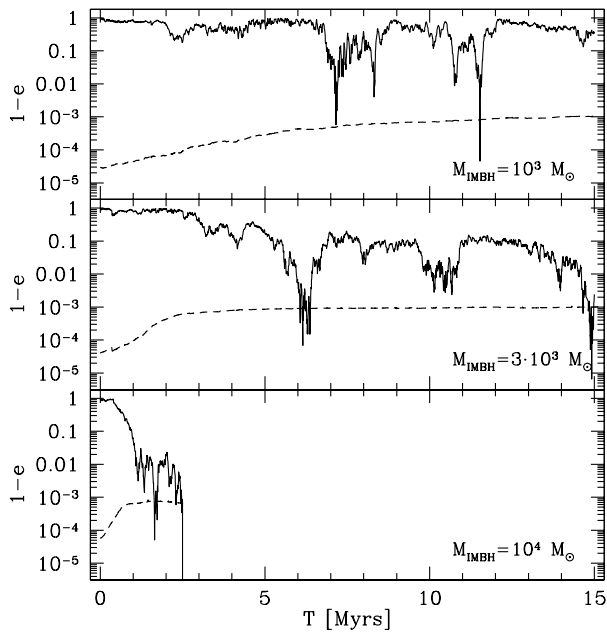}
\end{center}
\caption{Evolution of the orbital eccentricity of the 
inspiralling IMBHs. The orbits remain nearly circular 
during the inspiral phase and become highly eccentric 
in the stalling phase. The dashed lines show the eccentricity at which the
inspiral time becomes smaller than 10 orbital times. The IMBHs reach such
high-eccentricity orbits within a few Myrs, implying that the lifetime of
an IMBH in the galactic centre is very short.}
\label{fig:ecc}
\end{figure}
Fig.\,\ref{fig:ecc} depicts the evolution of orbital eccentricity of
the IMBHs. The orbital eccentricity of an IMBH is influenced by
dynamical friction, which arises due to many distant encounters, and a
few close interactions.  During the inspiral phase, the orbits of all
IMBHs stay nearly circular, despite a decrease in semi-major axis by
almost a factor of 100. This is in agreement with analytic estimates
which predict that dynamical friction in power-law halos with an
isotropic velocity distribution should circularise the orbit of an
infalling body \citep{ts00}.

After the IMBHs have reached the centre and removed most stars from
the inner cusp, dynamical friction is unimportant and the orbits of
the IMBHs change due to interactions with passing stars coming from
larger radii. In this phase, the orbits acquire eccentricities as high
as $1-e=10^{-5}$ or larger, which, in case of the $M_{\rm IMBH}=10^4$
M$_\odot$ IMBH, was high enough for the IMBH to pass within the radius
of the last stable orbit around the SMBH, resulting in the merger of the
two black holes.

In the absence of perturbations, the time for two orbiting black holes
to merge due to the emission of gravitational waves can be approximated with
(Peters 1964):
\begin{equation}
	T_{GW} = 0.0353 \; \frac{a^4 c^5}{G^3 m_1 m_2 (m_1+m_2)} 
		 \left( 1-e^2 \right)^{7/2}\,.
\label{eqgr}
\end{equation}
Here, $a$ and $e$ are the semi-major axis and eccentricity of the
orbit of the two black holes, $m_1$ and $m_2$ are their masses and $c$
is the speed of light.  The dashed lines in Fig.\,\ref{fig:ecc} show
the eccentricity for which the time for merging due to GW emission becomes
smaller than 10 orbital periods.  In this regime the emission of
gravitational waves dominates the evolution of the orbit, and the two
black holes are likely to merge before dynamical interactions can
reduce the eccentricity again.  It can be seen that the lifetime of an
IMBH in the galactic centre is limited to a few Myrs.  The time over
which stars are ejected by an IMBH is therefore also limited to a
few Myrs, i.e; the majority of HVSs are generated in short bursts.
This sets IMBH induced ejections apart from encounters of stellar
binaries with a SMBH, which would create escapers nearly continuously
and is therefore an important criterion to distinguish between the two
cases.
\begin{figure}
\begin{center}
\includegraphics[width=8.3cm]{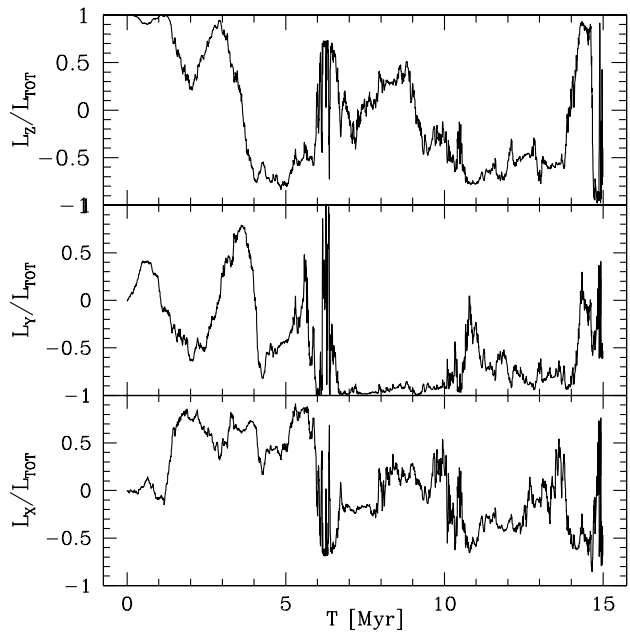}
\end{center}
\caption{Direction of the orbital angular momentum of 
the $M=3\cdot 10^3$ M$_\odot$ IMBH. 
Shown are the x (bottom panel), y (middle panel) and z component (top panel) 
of the angular momentum vector relative to the total angular momentum.
Initially, all IMBHs moved in the x-y plane. 
After the $M=3\cdot 10^3$ M$_\odot$ IMBH has spiralled into the centre,
the angular momentum vector changes its direction on 
a timescale of a Myr or less. The direction of motion of ejected  
stars will therefore also show correlations only on small timescales
while the general distribution should be nearly isotropic.}
\label{fig:angm}
\end{figure}

As explained earlier, another possibility to distinguish the ejection 
of HVS due to IMBHs from binary induced ejections is by
studying the spatial distribution of ejected stars. 
Since stars ejected by an IMBH acquire their
velocities mainly in the direction of motion of the IMBH, 
they should be ejected preferentially in the orbital
plane of the IMBH. If the orbit is eccentric during the inspiral phase, 
stars should be ejected mainly in one direction 
since the density of stars increases strongly towards the centre \citep{l05}.
In both cases it is however necessary that the orbital angular momentum vector
and the
Runge-Lenz vector of the IMBH stay constant for a sufficiently 
long time interval.

Fig.\,\ref{fig:angm} depicts the direction of the orbital angular 
momentum vector of the $M_{\rm IMBH}=3\cdot10^3$ M$_\odot$ IMBH as a function of time. Initially 
the IMBH orbits in the x-y plane, so $\vec{L}$ points towards the z-direction. 
For the $M_{\rm IMBH}=3\cdot10^3$~M$_\odot$ IMBH, the orientation of angular momentum stays 
approximately constant during the inspiral phase, but changes rapidly 
once the IMBH has reached the inner cusp and undergoes frequent interactions 
with cusp stars. After reaching the centre, the angular momentum vector changes 
on timescales of 1\,Myr, 
and even more rapidly when the IMBH moves on a high-eccentricity orbit. 
Any correlation between the escape direction of different HVS is therefore
destroyed on a timescale of a few Myrs and the general distribution 
should be nearly isotropic. We obtain this behaviour for all three 
IMBH masses and we will come back to this point when discussing the spatial distribution of
HVS in $\S$\,\ref{sec:hvs:distr}.
\begin{figure}
\begin{center}
\includegraphics[width=8.3cm]{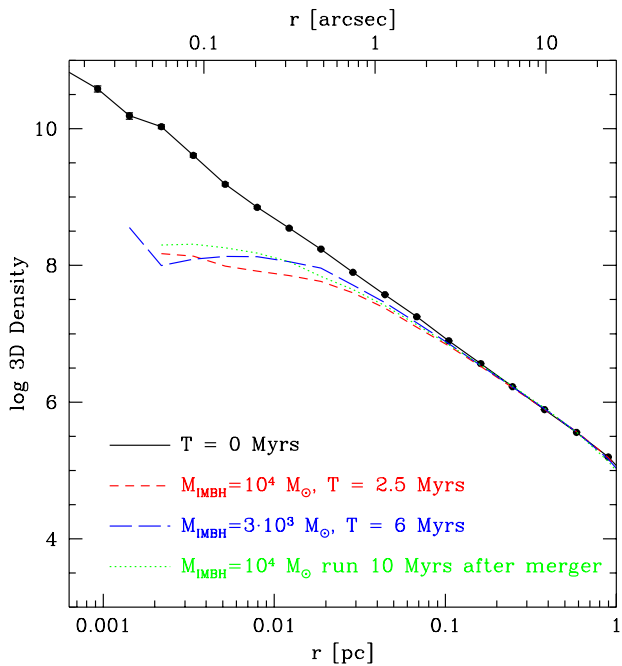}
\end{center}
\caption{Density distribution of stars at three different times 
for the $M_{\rm IMBH}=10^4\msun$ and $M_{\rm IMBH}=3 \cdot 10^3\msun$ runs.
In both runs, the initial density distribution follows a 
$\rho \sim r^{-1.75}$ power-law cusp which is depleted in the central parts 
and turned into a core profile after the IMBHs have spiralled into 
the centre. After the $M_{\rm IMBH}=10^4\msun$ merged with the SMBH, the 
central cusp is replenished only very slowly (dotted lines). 
The density profile in the Galactic centre should show 
a similar core if it contained an IMBH within the last 100\,Myrs.}
\label{fig:dens}
\end{figure}

Fig.\,\ref{fig:dens} shows the density profile of the stellar cusp 
for the $M_{\rm IMBH}=10^4\msun$ and $M_{\rm IMBH}=3\times 10^3\msun$
runs at the start and by the time the runs were stopped.
The initial density profile follows an $\alpha=1.75$ power-law 
cusp for radii $r<1$\,pc down to about $r=3 \times 10^{-4}$\,pc, 
at which radius the cusp runs out of stars. 
Due to the merger of cusp stars with the central SMBH and the ejection 
of stars by the IMBHs, the initial cusps are turned into 
core profiles with a core radius of about $r=0.02$\,pc. 
The core radii stay nearly constant in time since 
the ejection rate of stars is low after the initial peak. 
The only exception is the $M_{\rm IMBH}=10^3\msun$ run which did not 
last long enough to completely deplete the central cusp.

Although so far no indication for a cored density profile has been 
found in the Galactic centre, present observational techniques 
are only about now reaching the spatial resolution to measure the 
density profile inside 1''. If future observations reveal a cored 
density profile this would be additional evidence for an IMBH 
in the Galactic centre. In order to test how quickly the core 
is replenished, we continued the $M_{\rm IMBH}=10^4\msun$ run after
the IMBH merged with the SMBH. Even after running the simulation for 10\,Myrs, there was only a slight 
increase in central density. Since the relaxation time in our runs is about 
a factor of 10 smaller than in real Galactic nuclei, replenishing the 
cusp in the Galactic centre should take at least $T \approx 100$\,Myrs 
and possibly even longer. Similar large refilling times were also obtained by 
Wang \& Merritt (2005) through analytic estimates.
If observations show that the power-law cusp in the Galactic centre 
extends down to radii much smaller than 0.02\,pc, 
the presence of massive IMBHs in the Galactic centre within the 
last 100 Myrs could be excluded. In this case the observed HVS 
would not have been ejected by IMBHs, since typical travel 
times of HVS are of order $10^8$ years or less \citep{b06a}.

\subsection{Ejection of HVS}

As the IMBHs sink towards the Galactic centre, 
interactions with the background stars that orbit the SMBH result 
in a large number of ejections. 
We consider as escaping stars all stars that acquire positive energies 
during the calculation, i.e. which can leave the inner cusp region modelled 
in our simulations.
\begin{figure}
\begin{center}
\includegraphics[width=8.3cm]{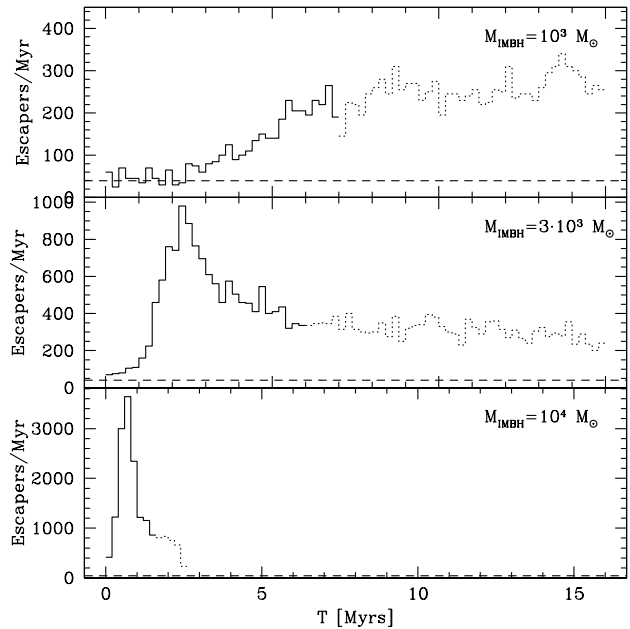}
\end{center}
\caption{Escape rate of stars ejected from the central cusp 
as a function of time (solid and dotted lines). 
The stellar escape rate rises once 
the IMBHs have reached the centre and drops as the IMBHs deplete
the central cusps. The dashed lines show the contribution 
of star-star interactions to the escape rate, which are responsible 
for only a small fraction of the escapers. 
Dotted lines show the part of the inspiral
which would not have been reached if we had included GW emission.}
\label{fig:escr}
\end{figure}
Fig.\,\ref{fig:escr} depicts the escape rate of all ejected stars 
as a function of time. The overall evolution is very similar for the
three IMBH masses: The escape rate rises with time and reaches a maximum 
when the IMBHs have spiralled into the centre. It drops on a similar timescale 
during which the IMBHs scatter away all stars from the inner cusp region 
(see also Fig.\,\ref{fig:dens}). At later times, escapers are created 
mainly by relaxation processes in the outer cusp, due to which stars with 
large semi-major axes can be scattered to low angular momentum orbits 
and reach the inner cusp. 

As shown in Fig.\,\ref{fig:ecc}, the orbital evolution of the IMBH
around the SMBH may be terminated by the emission of gravitational
waves before the end of the calculation. We have therefore shown the further evolution of the
escape rate in Fig.\,\ref{fig:escr} with the dotted lines.
For the higher mass IMBHs ($M_{\rm IMBH} \ge 3\times 10^3\msun$),
the majority of escapers are ejected well before the two black holes
merge due to GW emission.  

The dashed lines in Fig.\,\ref{fig:escr} show the contribution 
of star-star interactions to the escape rate. 
Since we did not check for collisions between the stars 
during the simulations, the rate of star-star escapers 
should be higher in our runs than it would be in reality.
We can estimate the rate of star-star escapers from the 
$M_{\rm IMBH}=10^3\msun$ run where in the beginning the IMBH 
moves through a low-density environment and encounters only few stars. 
We find about 40 escapers/Myr in this phase, which gives an upper limit 
for the rate of star-star escapers. We find a similar escape rate 
for the $M_{\rm IMBH}=10^4\msun$ run after the IMBH merged with the SMBH. 
Hence, star-star interactions contribute only a small fraction 
to the overall escape rate once the IMBHs have reached the centres 
and can therefore be neglected when discussing the properties of HVS.

In order to find the HVSs in our simulations, we followed the orbits 
of all escapers in the potential of the Galaxy after they left 
the Galactic centre.
We chose a fifth order Runge-Kutta method 
with adaptive step-size as the integrator.
The model for the Galactic potential 
was a combination of four separate components:
the galactic centre, represented by a power law density profile \citep{gen03},
the bulge, represented by a Plummer model, 
the disc, represented by a Kuzmin axisymmetric profile 
and the halo, represented by the Paczynski model \citep{p90}.
We computed the trajectories of all escapers for 200\,Myr or until
the stars reached a distance of 100\,kpc from the centre and 
considered as HVSs only those stars which acquired large enough velocities 
to be unbound and escape the Milky Way potential.  

\begin{figure}
\begin{center}
\includegraphics[width=8.3cm]{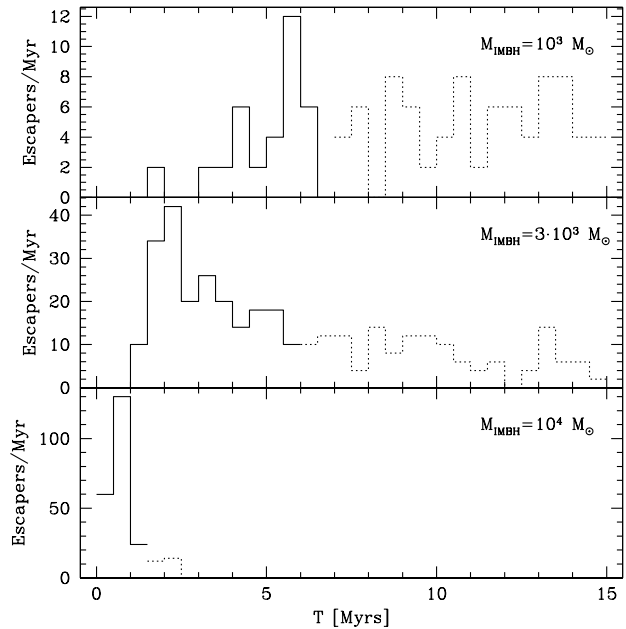}
\end{center}
\caption{Number of HVS created per Myr as a function of time 
for different IMBH masses. The escape rate of HVSs shows 
a similar trend to the overall escape rate and peaks when the
IMBHs arrive in the centre.}
\label{fig:hvs_esc}
\end{figure}

Fig.\,\ref{fig:hvs_esc} shows the rate of HVS ejected from the runs. 
The escape rate of HVSs follows a trend very similar to the overall 
escape rate. It reaches a sharp maximum when the IMBHs reach the centre 
and drops as the central cusps are depleted in stars. 

The largest escape rates, which are reached when the IMBHs have spiralled 
down to a radius of about $r=0.002$\,pc are 150 stars/Myr 
for $M_{\rm IMBH}=10^4\msun$, 42 stars/Myr for $M_{\rm IMBH}=3\cdot 10^3\msun$ 
and 15 stars/Myr for $M_{\rm IMBH}=10^3\msun$, scaling roughly 
with $M_{\rm IMBH}$. These numbers could be a factor of 10 to 30 higher 
for real Galactic nuclei, depending on the central density of stars. The 
average rate of HVS after the cusps have been depleted are 15 stars/Myr 
for $M_{\rm IMBH}=10^4\msun$,
10 stars/Myr for $M_{\rm IMBH}=3\cdot 10^3\msun$ 
and 4 stars/Myr for $M_{\rm IMBH}=10^3\msun$, scaling roughly 
with $M_{\rm IMBH}^{1/2}$. As explained in section 2.1, 
we expect these numbers to be independent of the average  mass of stars in our runs. 
As already explained before, an IMBH creates HVS only in a short span of time
before merging with the central SMBH. However, even if the IMBH can somehow avoid merging, 
HVS will show a peak in their escape times since
depending on which IMBH mass we assume, about as many stars escape within 1\,Myr after
the IMBH has reached the centre as would escape within
the following 40 to 100\,Myrs.

Based on the above numbers, and assuming that the IMBH merges with the SMBH
after 5\,Myrs, we can estimate that IMBHs with masses 
of $10^4\msun$, $3 \cdot 10^3\msun$ or  $10^3\msun$ 
create 1700, 900 and 600 HVS respectively. If we assume a standard IMF,
about 3\% of these stars would be O or B type stars. Especially
for the high-mass IMBHs, the numbers are therefore large enough 
to explain the 6 observed HVS.

\begin{figure*}
\begin{center}
\includegraphics[width=8.3cm]{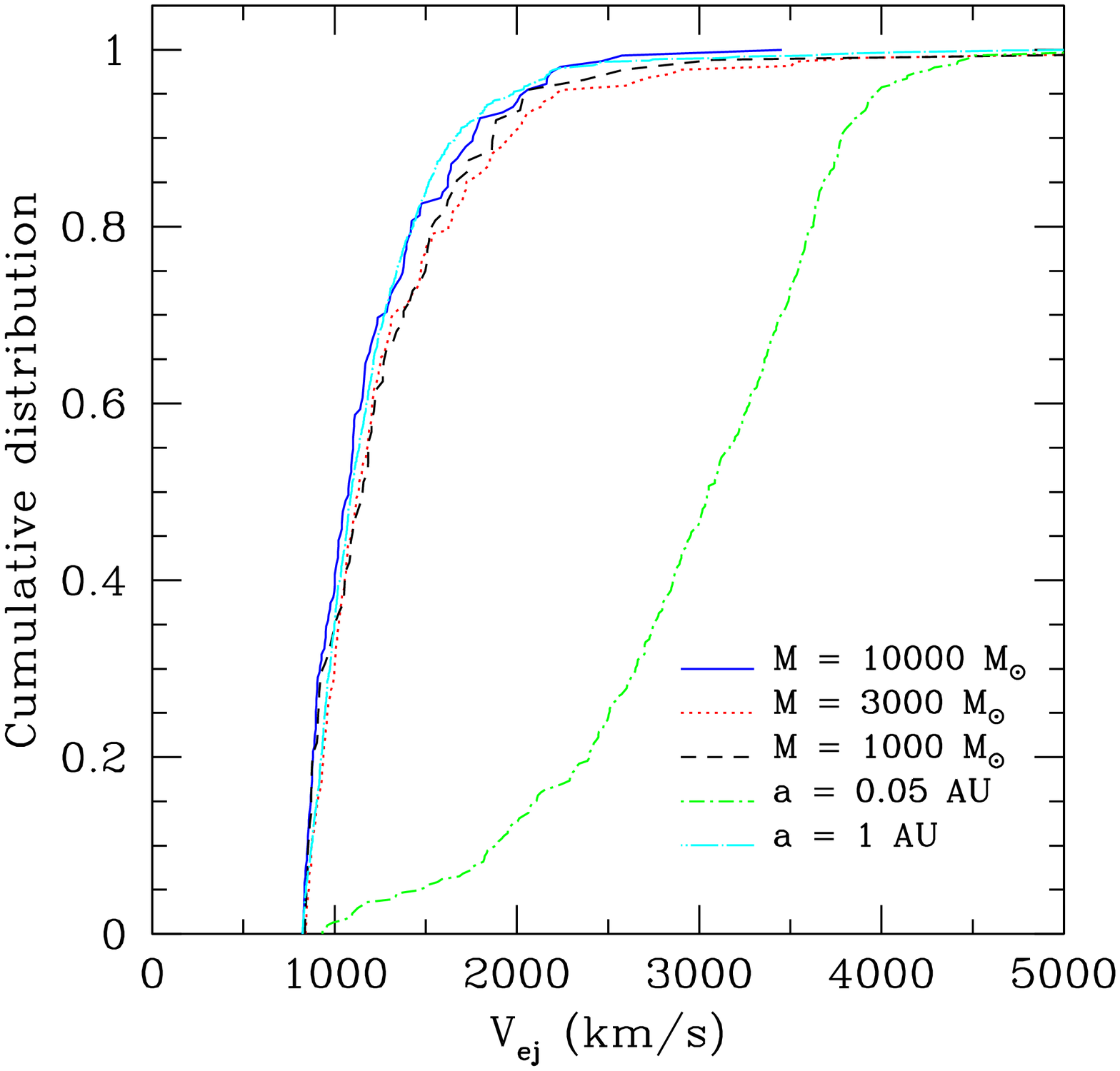}
\includegraphics[width=8.3cm]{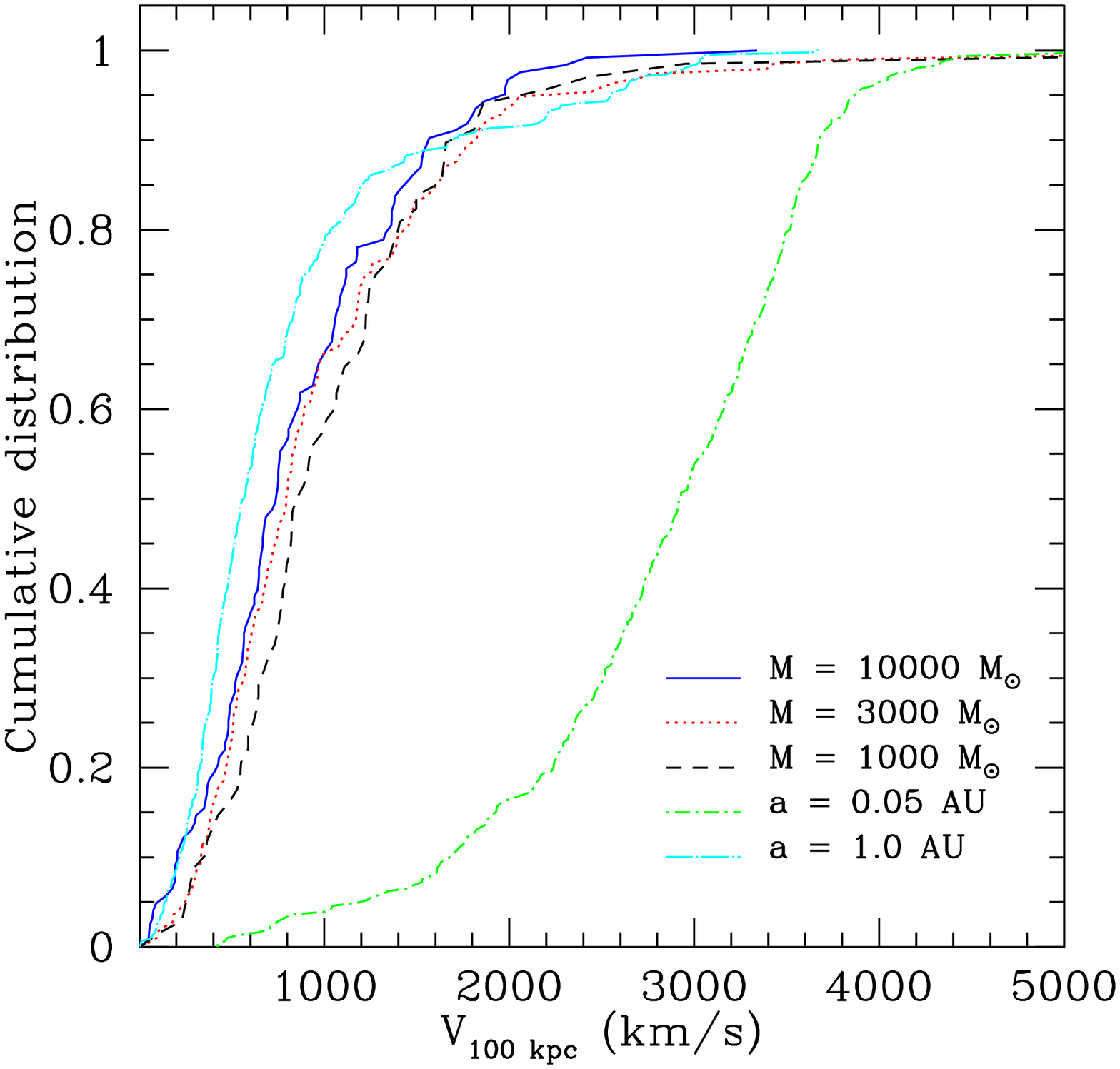}
\end{center}
\caption{Velocity distribution of HVS for three different
black hole masses (solid line, dotted line, dashed line), compared
with the distribution of escapers generated by encounters
between stellar binaries and the SMBH (dash-dotted line and long
dash-dotted line, see \citet{gps05}). Left panel shows the velocity distribution
of HVS when they are ejected from the cusp, right panel shows the distribution
after they have reached a distance of 100 kpc from the galactic centre.
The HVS distribution created by stellar binaries can be significantly different from 
the one created by IMBHs, which is nearly independent of the IMBH mass.}
\label{fig:veld}
\end{figure*}
Fig.\,\ref{fig:veld} shows the cumulative distribution of ejection velocities 
of HVS for the three different IMBH runs (solid line, dotted line, dashed line). 
The left panel shows the distribution when the HVS are ejected from the cusp, while the
right panel shows the distribution when the HVS have reached
a distance of $R=100$ kpc from the galactic centre. The distribution was determined 
from all escaping HVSs, but introducing a cut off in time due to the possible merger of the
IMBHs has little effect.
At ejection, the velocities span a range between $\sim800\kms$ and $\sim5000\kms$,
with the minimum being determined by the Galactic escape velocity. 
The distribution is very similar for the three runs, 
implying that the velocity with which stars escape does not depend 
strongly on the IMBH mass but rather on the local orbital velocity 
at the radius where the interaction takes place.
We compare these distributions with those generated by encounters
between stellar binaries and the SMBH as simulated by \citet{gps05}.
We consider equal mass binaries with $3\msun$ stars 
and two extreme values for the initial semi-major axis: 
$a = 0.05$\,AU (dash-dotted line) and $a=1$\,AU (long dash-dotted line).
The distribution for the smallest semi-major axis, which is the one
that generates the fastest escapers, is significantly different
from the distribution of HVS ejected by an IMBH.
When a large sample of HVSs will be available, 
it will be possible to discriminate between the two ejection
scenarios based on the measured velocity distribution.
We notice here that three-dimensional velocities are needed for this
kind of analysis.

\subsection{Spatial distribution of HVS}
\label{sec:hvs:distr}

\begin{figure*}
\begin{center}
\includegraphics[width=7cm]{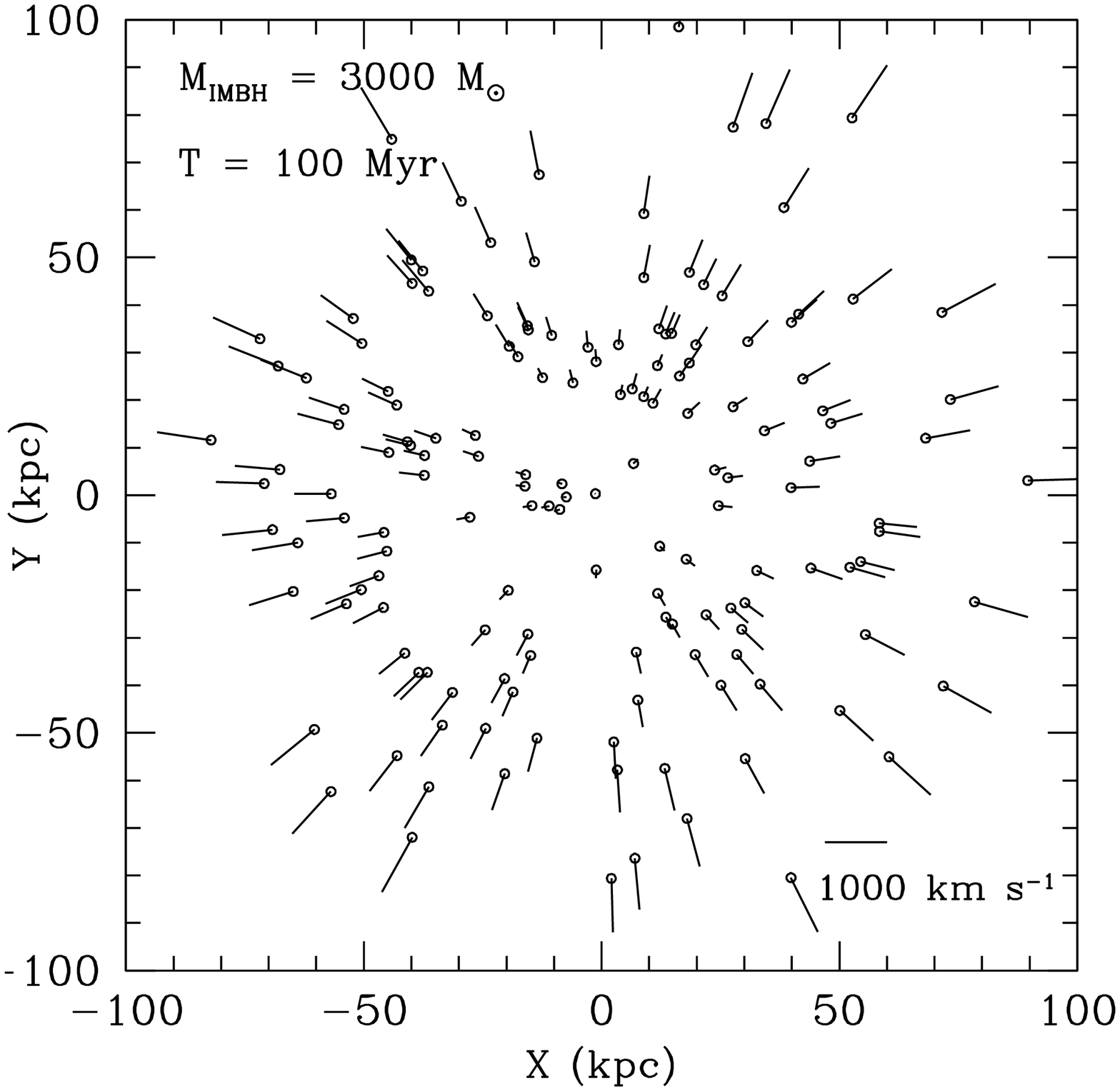}
\includegraphics[width=7cm]{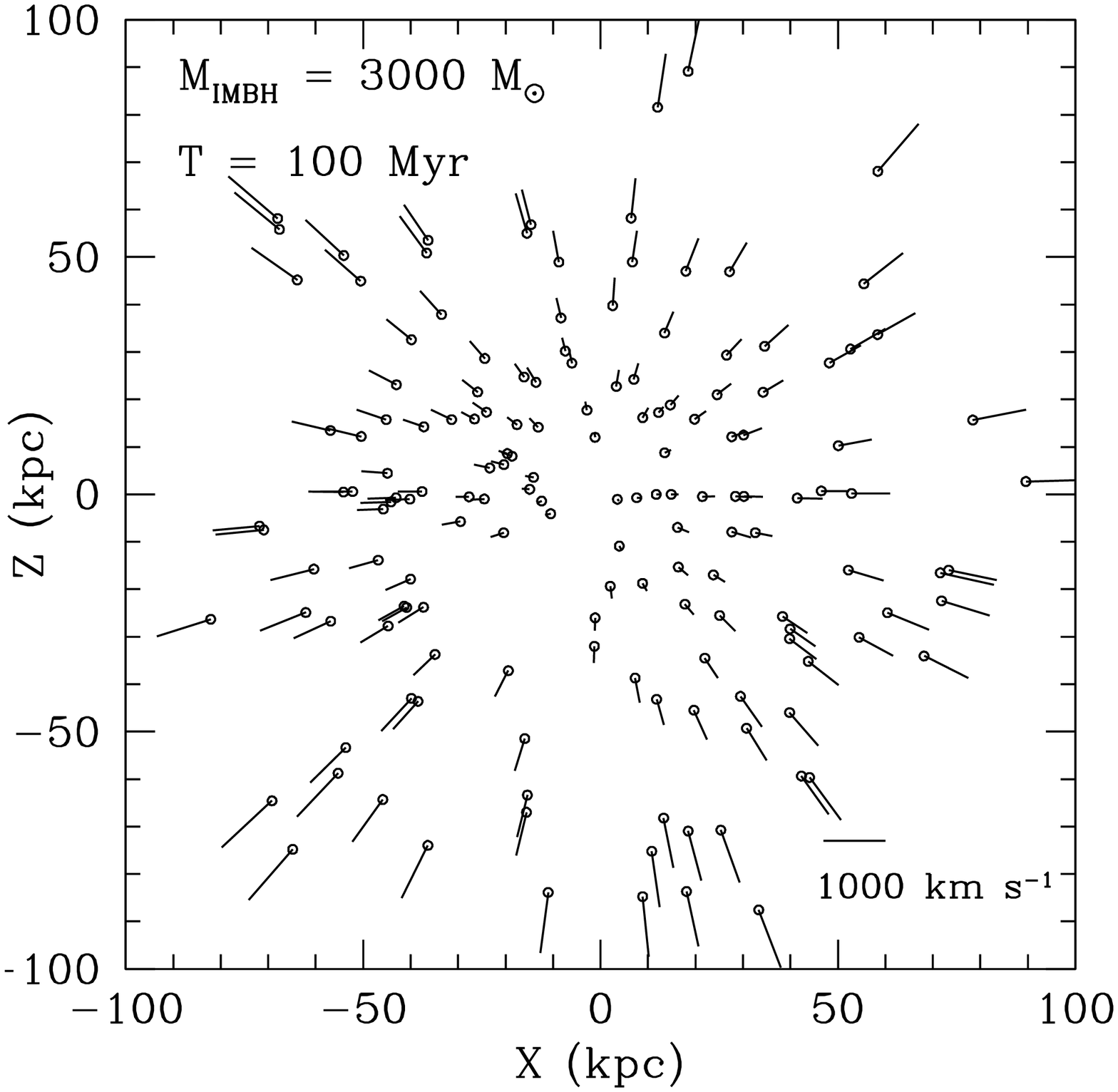}
\includegraphics[width=7cm]{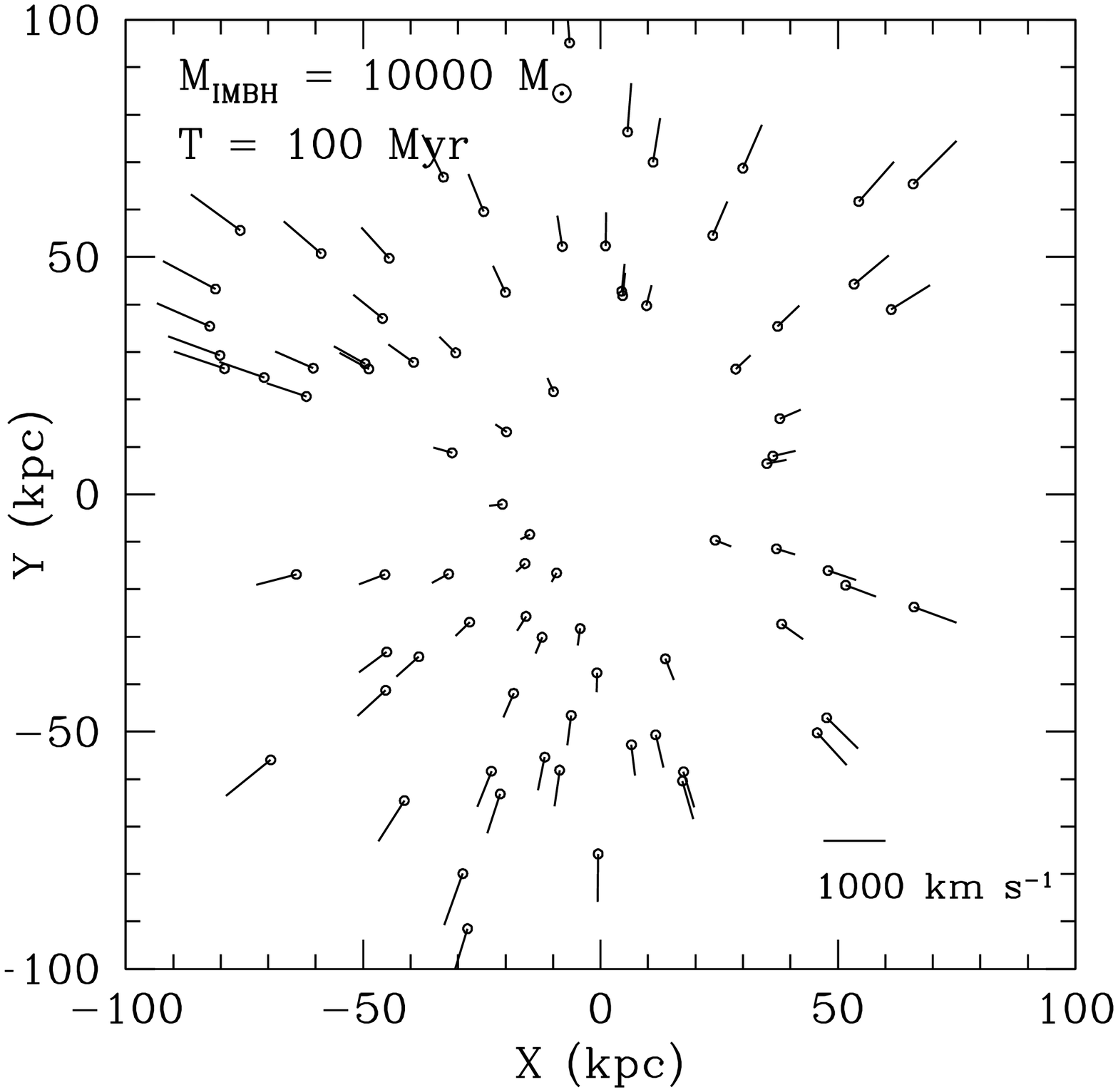}
\includegraphics[width=7cm]{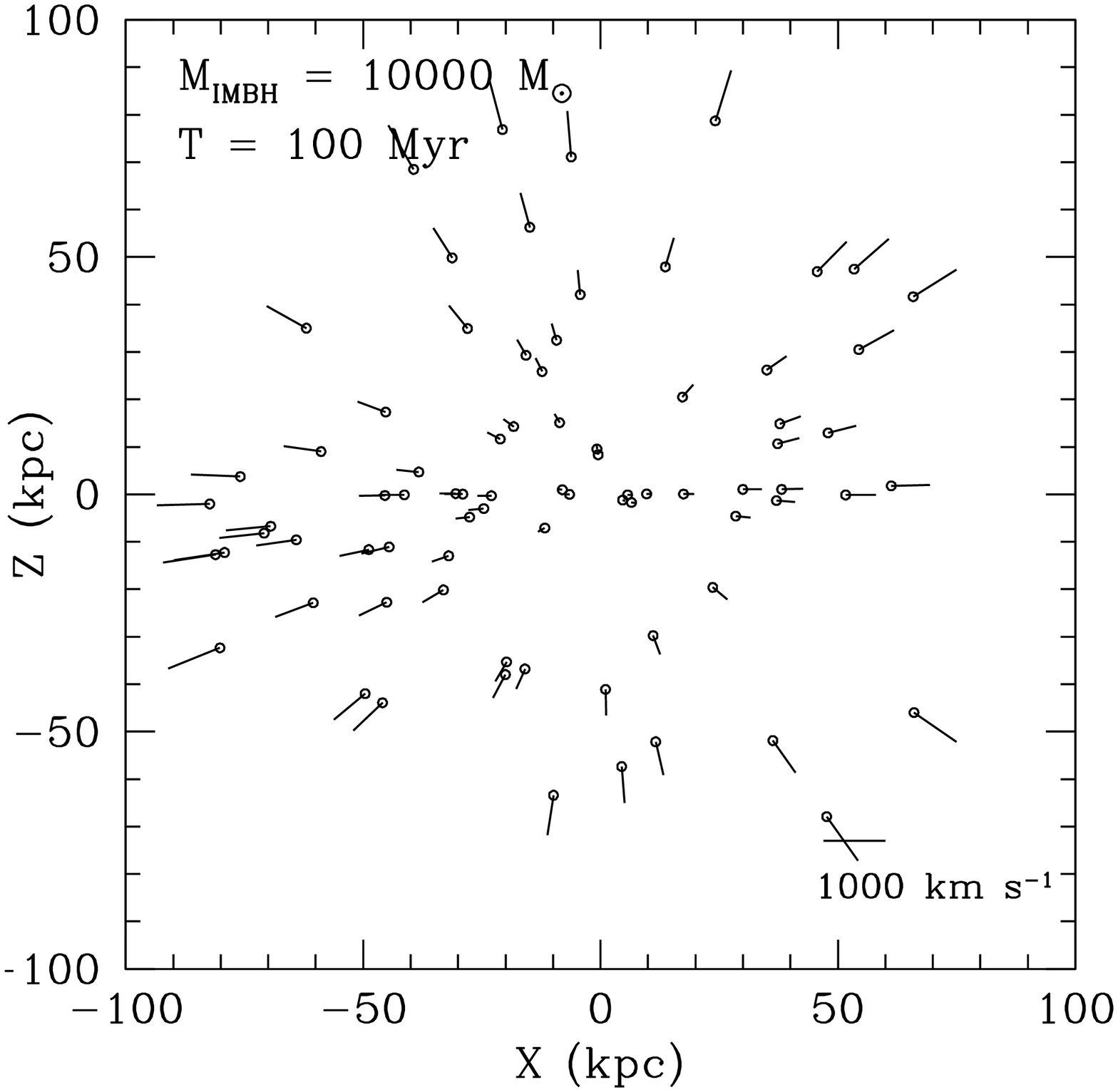}
\end{center}
\caption{Spatial distribution of all HVS created in the
$M_{\rm IMBH}=3000\msun$ run (upper plots) and in the
$M_{\rm IMBH}=10^4\msun$ run (lower plots) 
100\,Myrs after the start of the simulations.
The orbits of all escapers are followed in a realistic 
3D Galactic potential to obtain their spatial distribution.
The Galactic disc is assumed to be in the x-y plane.
The distribution of HVS after 100\,Myrs is isotropic in the x-y plane
for both IMBHs, but some overdensity
can be seen in the $M_{\rm IMBH}=10^4\msun$ case around z=0.
Also shown are the velocity vectors of each star. If HVS are
created by IMBHs, the fastest stars should be found at the largest 
distances.}
\label{fig:orb}
\end{figure*}
Fig.\,\ref{fig:orb} shows the spatial distribution of HVS 
for the $M_{\rm IMBH}=3000\msun$ and the $M_{\rm IMBH}=10^4\msun$ runs 
100\,Myrs after the start of the simulation
during which the orbits of the HVS have been followed 
in the potential of the Galaxy, assuming that the inspiral plane of the
IMBH agrees with the plane of the galactic disc (z=0).
For the $M_{\rm IMBH}=3000\msun$ IMBH, the distribution of HVS is nearly isotropic 
and would thus be indistinguishable from a HVS distribution created by encounters of
stellar binaries with a single SMBH. 
For the $M_{\rm IMBH}=10^4\msun$ case, there is an overdensity of stars
around the z=0 plane. This is due to the fact that the inspiralling 
IMBH had $L_z/\sqrt{L_x^2+L_y^2+L_z^2} \approx 1$ during the first 
Myr of the inspiral, and consequently ejected stars preferentially in the x-y plane. 
For the other runs we could not find such overdensities 
and they disappear for the 
$M_{\rm IMBH}=10^4\msun$ IMBH if we assume that 
the orbital plane of the IMBH does not coincide with the Galactic plane.
Although it is not impossible to find a non-isotropic spatial 
distribution of HVS, we conclude that this is at least unlikely.

Also shown in Fig.\,\ref{fig:orb}  are the velocity vectors of the HVS. 
As expected, they all point away from the galactic centre.
Since HVS are created in a short burst whose duration is much smaller 
than the time required to travel into the galactic halo, the escape
velocities of HVS also increase in magnitude with the distance from the centre.
In case HVS are created by an IMBH, we therefore expect a similar 
correlation between escape velocity and galactocentric distance 
with the fastest HVS being found at the largest distances.

\subsection{Dependence on the assumed mass of stars}
\label{sec:scal}

As explained in section 2.1, the average mass of stars in our runs is
probably too high compared to real galactic nuclei. In order to test which
influence this might have on our results, we repeated the $M_{IMBH} = 3000 M_\odot$
run, using stars with higher masses. In the first case we used $N=5 \cdot 10^4$
stars with average mass $m=60 M_\odot$ and in the second case $N=2.5 \cdot 10^4$ stars with
$m=120 M_\odot$. All other parameters, like the density distribution of the stars 
and the total mass in stars were held constant. 

\begin{figure}
\begin{center}
\includegraphics[width=8.3cm]{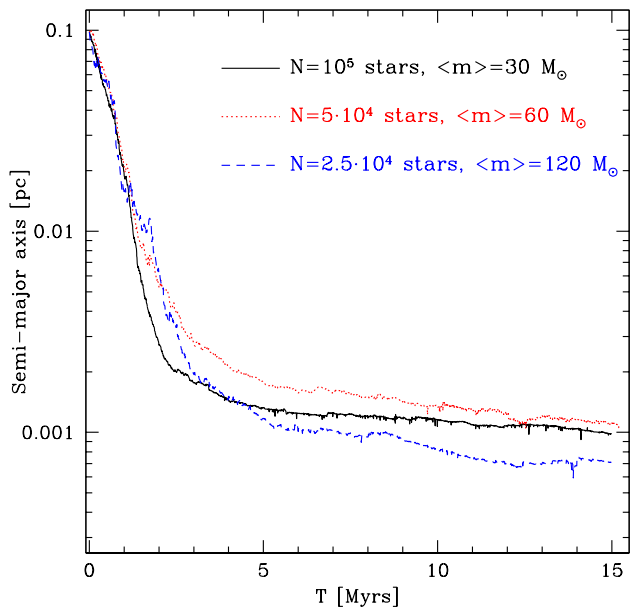}
\end{center}
\caption{Semi-major axes as a function of time of three IMBHs with mass $M_{IMBH}=3000 M_\odot$ 
moving through nuclei composed of stars with different stellar masses $m$. The resulting inspiral 
is very similar for all three IMBHs, showing that the inspiral results do not depend on 
the average mass of stars used in our runs.}
\label{fig:insp_scal}
\end{figure}

\begin{figure}
\begin{center}
\includegraphics[width=8.3cm]{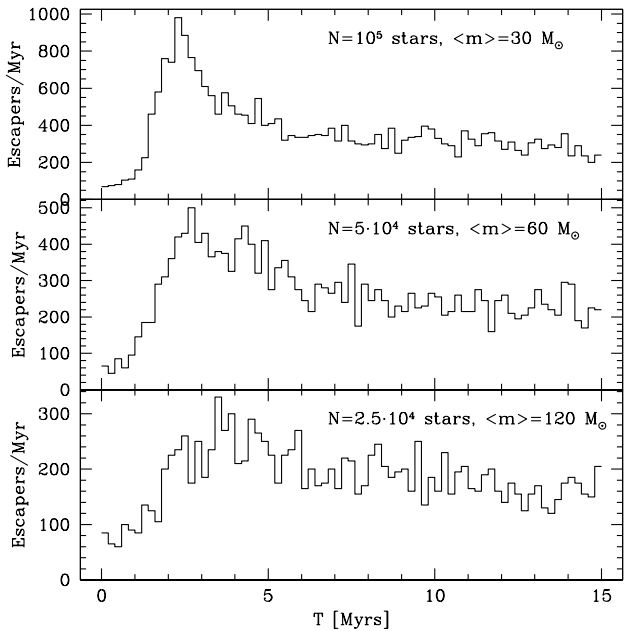}
\end{center}
\caption{Rate of stars ejected from the central cusp
as a function of time for an IMBH mass of 3000 $M_\odot$ but three different 
stellar masses. As expected
from theoretical arguments, the maximum ejection rate drops linearly
with the number of stars while the escape rate at later times is nearly
independent of the average stellar mass.}
\label{fig:eject_scal}
\end{figure}

Figs.\ \ref{fig:insp_scal}
and \ref{fig:eject_scal} show the results for the inspiral of the IMBHs and the
ejection of hypervelocity stars. Since $M_{IMBH} >> m$, the inspiral of the IMBHs should not
depend on the mass of individual stars but only on their overall density and this
is confirmed by Fig.\ \ref{fig:insp_scal} since all three black holes spiral in at more or
less the same rate. The slight differences visible in 
Fig.\ \ref{fig:insp_scal} between different runs are probably due to statistical 
effects since they become important only in the innermost part of the cusp where only
relatively few stars are present and do not show 
a trend with the stellar mass used. 
We also find that the eccentricity evolution of the IMBHs is independent of the stellar 
mass:
In all 3 runs, the orbits of the IMBHs stay nearly circular in the inspiral
phase and become highly eccentric in the stalling phase. All IMBHs would merge with the
SMBH after a few Myrs due to gravitational wave emission from high-eccentricity orbits.

Fig.\ \ref{fig:eject_scal} depicts the ejection rate of stars in the three runs.
As discussed in sec. 2.1,
we expect the maximum ejection rate to drop linearly with the number of stars in the cusp
and this is confirmed by our runs since we obtain maximum ejection rates of N=1000, 500 
and 300 stars/Myr for runs with $N=10^5, 5\cdot10^4$ and $2.5\cdot10^4$ stars. 
Fig.\ \ref{fig:eject_scal} also shows that in the later
phases, when the central cusp has been depleted in stars, the ejection rate is nearly 
independent of the number of stars which is also in very good agreement with our 
theoretical
considerations in 2.1. We finally found no significant dependence of stellar escape 
velocity with the mass of stars used, and conclude therefore that the results of our 
paper should be
robust against changes of the average stellar mass.

\section{Conclusions}
\label{sec:concl}

We have performed simulations of the inspiral of massive black holes
into the centres of galaxies and of the subsequent ejection of
hyper-velocity stars. We found that the spatial distribution of HVS is
nearly isotropic and would be difficult to distinguish from a HVS
distribution created by interactions of stellar binaries with an SMBH
if only few HVS were found, as is presently the case. A better
indication comes from the escape times of HVS: our simulations confirm
that most HVS are ejected in a short burst, lasting only a few Myrs
for typical IMBH masses, as soon as the IMBH reaches the galactic
centre. The ejection ends when the IMBH merges with the central SMBH,
which should take less than 10\,Myrs. Even if merging can be avoided,
the ejection rate of HVS is a factor of 30 to 100 lower than during
the burst maximum. The currently observed HVS show a broad
distribution of escape times \citep{b05}, which argues
against ejection due to a single IMBH, but would still be consistent with
the inspiral of several IMBHs.  The evidence is not conclusive
yet and requires more HVS to be found, which should be possible with
future astrometric surveys like e.g. {\it GAIA}. In case HVS are
ejected by an IMBH, we also expect a strong correlation of escape
velocity with galactocentric distance in the sense that the fastest
HVS can be found at the largest distances.

Another prediction from our runs is that IMBHs deplete the central
region in stars so that an initial cusp profile is turned into a
nearly constant density core with core radius $r=0.02$\,pc. If an IMBH
was present in the Galactic centre within the last 100 Myrs, such a
core should still be visible in the stellar density distribution.  If
on the other hand the cusp profile observed at larger radii continues
all the way down to the centre, this would be evidence against
the presence of an IMBH in the Galactic centre within the last $T
\approx 100$ Myrs.

\section*{Acknowledgements}
We are grateful to Sverre Aarseth for helping us with NBODY4, and to
Yuri Levin for discussions. HB thanks the University of Amsterdam for
their hospitality. This work was supported by the DFG Priority 
Program 1177 'Witnesses of Cosmic History: Formation and evolution 
of black holes, galaxies and their environment' and by the Netherlands
Organization for Scientific Research (NWO, \#635.000.001), the Royal
Netherlands Academy of Arts and Sciences (KNAW) and the Netherlands
Research School for Astronomy (NOVA).

\label{lastpage}

\end{document}